\def\arxivversion{1}
\ifdefined\arxivversion
  \documentclass[sigconf,nonacm,balance=false]{acmart}
  \newcommand{\singlefigwidth}{\linewidth}
\else
  \documentclass[manuscript,screen,review]{acmart}
  \acmJournal{TQC}
  \newcommand{\singlefigwidth}{0.5\linewidth}
\fi
\setcopyright{none}
\acmDOI{}
\acmYear{}
\acmVolume{}
\acmNumber{}
\acmArticle{}
\acmMonth{}

\usepackage{stmaryrd}
\usepackage{braket}
\usepackage{array}
\usepackage{verbatim}
\usepackage{blkarray}
\usepackage{bm}
\usepackage{algorithm2e}
\SetKwComment{Comment}{/* }{ */}
\SetKwInOut{Input}{Input}
\RestyleAlgo{ruled}
\ifdefined\arxivversion
  \newcommand{\TODO}[1]{}
  \newcommand{\FM}[1]{}
\else
  \newcommand{\TODO}[1]{\textcolor{red}{[TODO: #1]}}
  \newcommand{\FM}[1]{\textcolor{purple}{[FM: #1]}}
\fi

\begin{document}
\title{GPU-Based Parallel BP Decoding for Quantum LDPC Memory}

\author{Ming Wang}
\email{mwang42@ncsu.edu}
\affiliation{%
  \department{Department of Computer Science}
  \institution{North Carolina State University}
  \city{Raleigh}
  \state{North Carolina}
  \country{USA}}

\author{Yanbo Zhao}
\email{yzhao62@ncsu.edu}
\affiliation{%
  \department{Department of Computer Science}
  \institution{North Carolina State University}
  \city{Raleigh}
  \state{North Carolina}
  \country{USA}}

\author{Ang Li}
\email{angli16@uw.edu}
\affiliation{%
  \department{Department of Electrical and Computer Engineering}
  \institution{University of Washington}
  \city{Seattle}
  \state{Washington}
  \country{USA}}

\author{Frank Mueller}
\email{fmuelle@ncsu.edu}
\affiliation{%
  \department{Department of Computer Science}
  \institution{North Carolina State University}
  \city{Raleigh}
  \state{North Carolina}
  \country{USA}}
\renewcommand{\shortauthors}{Wang et al.}

\begin{abstract}
Quantum low-density parity-check (qLDPC) decoding requires reliable
corrections and low latency over long syndrome histories. We extend
BP-SF, an oscillation-guided belief-propagation (BP) decoder with
independent syndrome-flip trials, by introducing windowed and
streaming operations.  Our CUDA implementation combines
disordered-memory retries, sliding windows, correlation-aware graphs,
and layered scheduling. Circuit-level simulations test fixed-basis
memory and complete logical Pauli-frame recovery. On the
$\llbracket144,12,12\rrbracket$ bivariate bicycle code at physical
error rate $p=0.002$, parallel BP-SF reduces 99.9th-percentile call
latency by $13.15\times$ relative to an eight-chain Relay-BP GPU
baseline on 100,000 matched samples for which every tested method's
initial BP stage fails within 80 iterations. Windowed BP-SF
maintains a flat logical error rate and nearly constant per-round
compute time through 512 rounds, reaching a $17.1\times$ speedup over
monolithic decoding. In 512-round full-frame streaming decoding with
2.97-ms syndrome arrivals, layered BP-SF on a transformed joint graph
with 20 retries reduces 99.9th-percentile correction-release age from
$166.29$ to $36.25$ ms relative to the joint graph $D_{XYZ}$ with flooding updates and 100
retries. These results connect parallel BP post-processing to lower
decoding costs and earlier correction release in streaming quantum
memory.

\end{abstract}

\maketitle
\ifdefined\arxivversion
  \pagestyle{plain}
  \thispagestyle{plain}
\fi

\section{Introduction}
One of the central challenges in quantum computing is
to make fault-tolerant quantum computing (FTQC) practical. In FTQC, a
common practice for keeping quantum states alive is to encode the
states using a quantum error correction (QEC) code, where ancillary
qubits are measured repeatedly to extract error information from the
quantum state, a process called syndrome extraction. Therefore, on the
classical side, we need decoders that are fast enough to keep up with
syndrome extraction and as accurate as possible to suppress the error
rate. Considerable effort~\cite{9773217, 10764437,
  10.1145/3579371.3589037, 10.1145/3503222.3507707,
  10.1145/3695053.3731022} has therefore focused on decoder algorithms
and supporting hardware. Among the QEC codes, quantum low-density
parity-check (qLDPC) codes have attracted significant attention in
recent years due to their potential to encode more logical qubits than
surface codes while maintaining a low error
rate~\cite{hgp,panteleev2021degenerate,Bravyi_2024}. Their decoding
problem, however, remains challenging. Reliable operation requires
errors to be removed accurately and at a rate compatible with the QEC
cycle. In practice, the decoder must keep pace with syndrome
extraction to avoid a growing data backlog~\cite{RevModPhys.87.307},
making both algorithmic efficiency and implementation speed essential.

Belief propagation (BP) is an attractive starting point because its
sparse, and local message updates support efficient parallel
implementations, as established by classical LDPC
decoders~\cite{910578,1657127,su202258,5437474}.  For qLDPC codes,
however, degeneracy and trapping sets can leave BP oscillating between
ambiguous estimates~\cite{raveendran2021trapping}.  BP with ordered
statistics decoding (BP-OSD) addresses failed attempts through
algebraic post-processing~\cite{panteleev2021degenerate,roffe2020},
but the Gaussian-elimination stage adds computational cost. Localized
statistics decoding, ambiguity clustering, and stabilizer inactivation
reduce the scope of this algebraic work~\cite{hillmann2024localized,
  wolanski2024ambiguity,ducrest2022inactivation}. Alongside these
approaches, progress in BP-based decoding can be understood through
three complementary choices: how messages evolve, how the decoding
graph is represented, and how multiple decoding attempts are
organized.

The first choice concerns \emph{BP updates and
  scheduling}. Memory-based updates feed information from previous
iterations into the current beliefs~\cite{11027786}. Disordered memory
BP (DMem-BP) varies the memory strength across nodes to help break
symmetries that stall uniform updates~\cite{muller2025improved}.
Layered schedules change when new information becomes available to
neighboring nodes and can mitigate symmetric trapping
sets~\cite{raveendran2021trapping}. Other methods guide BP by freezing
selected variables or modifying their
priors~\cite{poulin2008iterative,10619083}. BP with bias using
oscillating trapping sets (BP-OTS), for example, identifies unreliable
variables from their oscillations and posterior beliefs, then assigns
a prescribed negative prior LLR~\cite{10409156}. This introduces a
bias-strength parameter in addition to the choice of variables. These
methods change the trajectory of an individual decoding attempt while
retaining message passing as the main computation.

The second choice is the \emph{graph representation}. Changing the
graph can remove structures that make local inference
unreliable. SymBreak splits selected parity checks to disrupt
degeneracy-induced symmetries~\cite{yin2024symbreak}, while ordered
Tanner forest decoding removes cycles before rerunning
BP~\cite{demarti2026otf}. Under circuit-level noise, the
representation also determines whether the decoder captures
correlations between $X$- and $Z$-type errors. Joint detector models
retain these correlations, and the GARI transformation rewires the
joint graph to eliminate four-cycles involving mixed-error mechanisms
while preserving the decoding problem~\cite{maan2025decoding}. Graph
design thus complements changes to the update rule, i.e., it changes
the dependencies on which those updates operate.

The third choice is \emph{trial diversity and ensembles}. Multiple
attempts can explore alternatives that a single BP trajectory
misses. Check-agnosia varies selected
priors~\cite{ducrest2024checkagnosia}, automorphism ensembles
transform the decoding input~\cite{koutsioumpas2025automorphism}, and
Relay-BP passes posterior information through successive DMem-BP runs
with different memory strengths~\cite{muller2025improved}. Related
search-based methods, including guided decimation guessing and
BP-guided beam search, explore multiple error hypotheses using BP
information~\cite{gong2024toward, ye2026beam}. The organization of
these attempts determines how much work can run concurrently and when
a result can be returned. The three choices are therefore not separate
algorithm families: GARI, for example, combines a transformed graph
with layered updates and an ensemble of randomized
schedules~\cite{maan2025decoding}.

\begin{figure*}[!t]
    \centering
    \includegraphics[width=\textwidth]{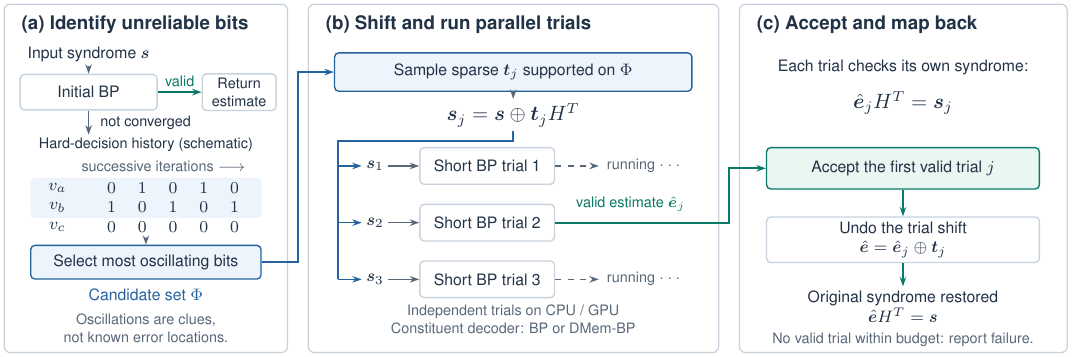}
    \caption{Overview of BP-SF. (a) An initial BP attempt returns
      immediately if its estimate satisfies the input syndrome;
      otherwise, hard-decision oscillations identify a candidate set
      $\Phi$. (b) Sparse trial vectors supported on $\Phi$ generate
      shifted syndromes for independent BP trials. (c) The first
      syndrome-valid trial is mapped back by undoing its trial
      vector.}
    \label{scheme}
\end{figure*}

Our previous work~\cite{wang2026bpsf} introduced BP with
syndrome flipping (BP-SF) within this ensemble approach. An
unsuccessful initial BP run identifies frequently oscillating
hard-decision bits. Sparse trial vectors supported on these locations
generate perturbed syndromes for independent BP attempts, and the
first syndrome-valid result is mapped back to the original problem
(Fig.~\ref{scheme}). The syndrome perturbation retains the channel
LLRs and avoids an additional prescribed bias magnitude. Once the
trial inputs are prepared, the attempts can execute concurrently
without Gaussian-elimination post-processing.

BP-SF separates the generation of trial inputs from the
message-passing algorithm used within each trial. This separation
gives us a way to bring recent improvements in BP updates and graph
representations into the same parallel decoding strategy. In this
article, we extend our prior work~\cite{wang2026bpsf} with a CUDA
implementation using posterior-initialized DMem-BP retries, then study
joint correlated-error models, the GARI transformation, and layered
scheduling within BP-SF. The question is how these choices affect the
accuracy--latency trade-off, including the long-running calls that can
delay subsequent corrections.

For quantum memory, that question extends beyond an individual
decoding call. Repeated syndrome extraction grows the decoding graph,
while waiting for more rounds delays the correction of earlier
errors. Sliding-window decoding bounds the active temporal region and
commits corrections
incrementally~\cite{huang2024sliding,gong2024toward}. FPGA
implementations of Relay-BP with windowing and GARI transformations
also demonstrate the importance of matching decoder organization to
the execution
platform~\cite{maurer2025realtime,bascones2026gari}. Here, we combine
windowing with GPU BP-SF and evaluate both fixed-basis memory and
recovery of the complete logical Pauli frame. We measure not only
decoding time, but also correction-release delay as syndrome data
arrive, so that the effects of lookahead and computational queuing
are included.

Building on the BP-SF ensemble first introduced in~\cite{wang2026bpsf}, we
make three additional contributions:

\textbf{GPU implementation with reduced tail latency:} We combine
BP-SF with posterior-initialized disordered-memory BP in a parallel
GPU implementation. This extends BP-SF beyond its original constituent
decoder while preserving independent trial execution. On the
$\llbracket 144,12,12\rrbracket$ code at \(p=0.002\), the
implementation reduces 99.9th-percentile call latency by
\(13.15\times\) relative to the tested eight-chain Relay-BP GPU
configuration on 100,000 matched samples for which every tested
method's native initial BP stage fails within 80 iterations.

\textbf{Scalability through windowed BP-SF:} We integrate BP-SF into a
sliding-window decoder and characterize how core and buffer lengths
balance accuracy, computation, and correction delay. On the 144-code
fixed-basis memory benchmark at \(p=0.002\), windowing mitigates the
convergence degradation of monolithic decoding, maintaining a logical
error rate near \(10^{-5}\) and nearly constant per-round compute time
through 512 rounds. At that length, it achieves a \(17.1\times\)
compute-time speedup.

\textbf{Full-frame correction:} We explore different constituent
decoding graphs on BP-SF and show that the choice of graph could
substantially reduce the trial budget and latency. For twelve-round
full-frame recovery, layered BP-SF on the GARI graph with 20 retries
achieves similar measured logical accuracy to the joint graph $D_{XYZ}$ with flooding updates
with 100 retries, while reducing mean latency by \(19.3\times\). In
a 512-round long memory experiment, these configurations yield
99.9th-percentile correction-release ages of 36.25 and 166.29 ms,
respectively.

\section{Background}
\subsection{Quantum LDPC Codes}

A stabilizer code specifies valid encoded states through a set of
commuting quantum parity checks, represented by Pauli operators. Each
check returns $+1$ on a valid encoded state while errors may change
some outcomes to $-1$. Recording these outcomes as binary values
produces the syndrome from which the decoder infers a correction. Let
$\mathcal{S}$ denote the group generated by these checks. The code
space $\mathcal{C}$ consists of states left unchanged by every element
of $\mathcal{S}$ :
\begin{equation}
    \mathcal{C}=\{\ket{\psi}\mid s\ket{\psi}=\ket{\psi},\ \forall s\in\mathcal{S}\}.
\end{equation}

An $\llbracket n, k, d\rrbracket $ stabilizer code encodes $k$ logical
qubits into $n$ physical qubits and has $n-k$ independent stabilizer
generators. The distance, $d$, is the minimum number of qubits on
which a Pauli error can act nontrivially while changing the encoded
information without changing any check outcomes. With ideal syndrome
measurements, the code can in principle correct any error affecting at
most $\lfloor(d-1)/2\rfloor$ qubits.

Calderbank–Shor–Steane (CSS) codes separate the stabilizer generators
into two types: checks containing only $X$ and identity operators, and
checks containing only $Z$ and identity operators. These checks are
represented by binary parity-check matrices $H_X$ and $H_Z$,
respectively. Each row represents a check, and each column represents
a physical qubit. A nonzero entry indicates that the check acts on
that qubit. For example, the row $(1,0,1,1)$ in $H_X$ represents the
check $X_1X_3X_4$. To ensure that all checks commute, every $X$-type
check must overlap every $Z$-type check on an even number of
qubits. In matrix form, this requirement is $H_XH_Z^T=0$, with
arithmetic performed modulo two. The two check types provide
complementary error information. An $X$ error reverses the outcomes of
the $Z$-type checks that involve the affected qubit, while a $Z$ error
reverses the corresponding $X$-type check outcomes. This structure
allows the two error components to be decoded separately using $H_Z$
and $H_X$, respectively. However, separate decoding need not capture
all available information because a $Y$ error contains both
components, creating correlations that a joint decoder can exploit.

For a CSS-type \emph{qLDPC} code family, both parity-check matrices
have bounded row and column weights as the code size
increases. Equivalently, each check involves only a bounded number of
qubits, and each qubit participates in only a bounded number of
checks. This limits the number of qubit--check interactions required
for syndrome extraction and gives the decoder a sparse representation
of the error constraints. The message-passing decoders considered in
this paper operate on this sparse structure.

\subsection{Syndrome Decoding and Belief Propagation}
The decoder receives a measured syndrome and prior information about
the errors that may have occurred. Let $H\in\mathbb{F}_2^{M\times N}$
describe $M$ parity constraints on $N$ binary error variables, and let
$\bm{s}\in\mathbb{F}_2^M$ be the syndrome. An error estimate,
$\hat{\bm{e}}$, is syndrome-consistent if $\hat{\bm{e}}H^T=\bm{s}$.
For CSS decoding, the variables can represent the $X$ or $Z$ error
components on physical qubits. Under circuit-level noise, the same
binary formulation instead uses error mechanisms as variables and
detector outcomes as constraints.

Syndrome consistency alone generally leaves many possible estimates.
For independent binary errors with prior probabilities, $p_i$, the
most likely individual error pattern is
\begin{equation}
\label{MLD_classical}
    \hat{\bm{e}}
    = \arg\min_{\bm{e}H^T=\bm{s}}
      \sum_{i=1}^{N} e_i \log\frac{1-p_i}{p_i}.
\end{equation}

Quantum correction does not require identifying the exact physical
error. A correction succeeds whenever its combination with the actual
error acts trivially on the encoded information. In particular, errors
that differ by a stabilizer have the same action on the code space and
belong to the same equivalence class. An optimal quantum decoder
therefore selects the most probable class: Given a syndrome, $\bm{s}$,
the optimal decoding problem considering degeneracy becomes
\begin{equation}
\label{MLD_quantum}
    [\hat{E}]
    = \arg\max_{[E]:\,\operatorname{synd}(E)=\bm{s}}
      \Pr([E]),
\end{equation}
where $[E]$ denotes the equivalence class of errors under the
stabilizer group and $\Pr([E])$ is the total probability of all errors
in that class. While the classical decoding problem in
Eq.~(\ref{MLD_classical}) is NP-hard~\cite{1055873}, its quantum
counterpart in Eq.~(\ref{MLD_quantum}) is even more complex, being
\#P-complete~\cite{10.1109/TIT.2015.2422294}. These complexity results
make exact decoding impractical at useful system sizes. Practical
decoders consequently approximate the optimum in a way that preserves
performance under the intended noise model.

BP provides such an approximation by exchanging messages on the Tanner
graph of $H$. A variable node, $v_i$, represents an error variable, a
check node, $c_j$, represents a parity constraint, and an edge
connects them when $H_{ji}=1$. We write $N(v_i)$ and $N(c_j)$ for
their respective neighbor sets. BP updates beliefs about individual
error variables rather than explicitly computing the probability of
every stabilizer equivalence class. Sparsity keeps BP updates
efficient, but short cycles in the graph deteriorate BP decoding by
producing circulating messages. What's more, symmetric trapping sets can
also prevent convergence~\cite{raveendran2021trapping}.

We use normalized min-sum (NMS), a low-complexity BP variant with
real-valued messages. A positive message favors an error variable
being zero, a negative message favors one, and its magnitude expresses
confidence. The decoder uses the following flooding schedule, in which
all messages of one type are updated before proceeding to the next type.

\begin{enumerate}

\item \textbf{Initialization:}
For a variable with prior error probability $p_{v_i}$, define the
channel log-likelihood ratio (LLR)
\begin{equation}
    l^{ch}_{v_i}=\log\frac{1-p_{v_i}}{p_{v_i}}.
\end{equation}
Initialize every variable-to-check message as
$l_{v_i\rightarrow c_j}^{(0)}=l^{ch}_{v_i}$.
The superscript $t$ denotes the iteration, starting at $t=1$.

\item \textbf{Check-to-variable update:}
Each check uses the messages from its other neighbors to determine
what the measured parity implies about a particular variable:
\begin{equation}
\label{c2v}
\begin{split}
    l_{c_j\rightarrow v_i}^{(t)}
    ={}&(-1)^{s_j}\alpha
       \min_{v_{i'}\in N(c_j)\setminus\{v_i\}}
       \left|l_{v_{i'}\rightarrow c_j}^{(t-1)}\right|\\
    &{}\times
       \prod_{v_{i'}\in N(c_j)\setminus\{v_i\}}
       \operatorname{sign}
       \left(l_{v_{i'}\rightarrow c_j}^{(t-1)}\right).
\end{split}
\end{equation}
The sign combines the neighboring decisions with the required syndrome
bit $s_j$. The magnitude is set by the least-confident contributing
neighbor and scaled by the normalization factor $\alpha$.
\item \textbf{Variable-to-check update:}
Each variable combines its channel LLR with the incoming check
messages, excluding the destination check:
\begin{equation}
\label{v2c}
    l_{v_i\rightarrow c_j}^{(t)}
    = l^{ch}_{v_i}
      + \sum_{c_{j'}\in N(v_i)\setminus\{c_j\}}
        l_{c_{j'}\rightarrow v_i}^{(t)}.
\end{equation}
Excluding the destination prevents its own message from being
immediately returned as new information.

\item \textbf{Decision and convergence test:}
The decoder combines all incoming check messages to form the soft
output
\begin{equation}
\label{eq:bp_marginal}
    l^{out,(t)}_{v_i}
    = l^{ch}_{v_i}
      + \sum_{c_{j'}\in N(v_i)}
        l_{c_{j'}\rightarrow v_i}^{(t)}.
\end{equation}
This output is an approximate belief and the hard decision is
\begin{equation}
    \hat{e}_i^{(t)}
    =
    \begin{cases}
        0, & l^{out,(t)}_{v_i}>0,\\
        1, & \text{otherwise}.
    \end{cases}
\end{equation}
The decoder returns the estimate if
$\hat{\bm{e}}^{(t)}H^T=\bm{s}$; otherwise, it continues until the
iteration budget is exhausted.
\end{enumerate}

Disordered memory BP (DMem-BP)~\cite{muller2025improved} replaces the
fixed channel LLR $l^{ch}_{v_i}$ in Eqs.~(\ref{v2c})
and~(\ref{eq:bp_marginal}) with a time-dependent LLR $l_{v_i}^{(t)}$.
At iteration $t$, this LLR mixes the channel LLR with the output LLR
from the previous iteration, weighted by a random memory strength,
$\gamma_{v_i}$. More specifically, it initializes
$l^{out,(0)}_{v_i}=l^{ch}_{v_i}$ and
$l_{v_i\rightarrow c_j}^{(0)}=l^{ch}_{v_i}$, and then updates for $t\geq 1$,
\begin{equation}
\label{eq:dmem_bias}
    l_{v_i}^{(t)}=(1-\gamma_{v_i})l^{ch}_{v_i}
        +\gamma_{v_i}l^{out,(t-1)}_{v_i}.
\end{equation}
Here, $l^{out,(t)}_{v_i}$ denotes the output LLR at iteration $t$.  The
memory strength $\gamma_{v_i}$ is sampled once for each variable node
and held fixed throughout a run. The output (marginal) LLR and
outgoing V2C messages become
\begin{equation}
\label{eq:dmem_messages}
\begin{aligned}
    l^{out,(t)}_{v_i}
        &=l_{v_i}^{(t)}+\sum_{c_{j'}\in N(v_i)}
            l_{c_{j'}\rightarrow v_i}^{(t)},\\
    l_{v_i\rightarrow c_j}^{(t)}
        &=l^{out,(t)}_{v_i}-l_{c_j\rightarrow v_i}^{(t)}.
\end{aligned}
\end{equation}

Setting every $\gamma_{v_i}=0$ recovers ordinary BP, while a uniform
$\gamma_{v_i}=\gamma$ with $0<\gamma<1$ gives Mem-BP, the binary
version of Chen et al.'s EWAInit-BP~\cite{11027786}.  Varying
$\gamma_{v_i}$ across nodes, including negative values, can break
symmetries that trap uniform memory
updates. Equation~(\ref{eq:dmem_bias}) shows that negative memory
reverses the feedback from the previous output LLR's deviation from
the channel LLR.

\subsection{GARI Transformation}
Under circuit-level noise, a single error mechanism can affect both
$X$- and $Z$-type detectors. Joint decoding captures the correlations
between these two sectors, but their shared connections can introduce
short cycles that hinder BP. Graph augmentation and rewiring for
inference (GARI)~\cite{maan2025decoding} addresses this problem by
introducing auxiliary variables and reorganizing the decoding graph.

For the joint detector model considered here, the decoding matrix is
\begin{equation}
D_{XYZ}=\begin{blockarray}{cccc}
\bm{e}_Z & \bm{e}_X & \bm{e}_Y \\
\begin{block}{(c|c|c)c}
  D_X    & \bm{0} & D'_X & \bm{s}_X  \\
  \bm{0} & D_Z    & D'_Z & \bm{s}_Z \\
\end{block}
\end{blockarray},
\end{equation}
Its columns correspond to error variables
$\bm{e}=(\bm{e}_Z,\bm{e}_X,\bm{e}_Y)$, and its rows correspond to
detector outcomes $\bm{s}=(\bm{s}_X,\bm{s}_Z)$, with
$\bm{e}D_{XYZ}^T=\bm{s}$.
The subscripts on $D_X$ and $D_Z$ identify the detector types:
$X$-type detectors respond to $Z$-type errors, and vice versa.
The primed blocks describe the contributions of mixed, or $Y$-type,
mechanisms to both sectors.

In this model, each column of $D'_X$ matches a column of $D_X$, and
similarly for $D'_Z$ and $D_Z$ because a $Y$-error can be decomposed
as an $X$- and $Z$-error on the same place.  Binary selection matrices
$U$ and $V$ record these matches, so that $D'_X=D_XU$ and $D'_Z=D_ZV$.
GARI introduces effective error variables
\begin{equation}
\label{eq:gari_effective}
\begin{aligned}
    \bar{\bm{e}}_Z &= \bm{e}_Z+\bm{e}_YU^T,\\
    \bar{\bm{e}}_X &= \bm{e}_X+\bm{e}_YV^T,
\end{aligned}
\end{equation}
These variables separate the contributions of $Y$-errors to $Z$- and
$X$-sector. The physical constraints therefore become
$\bar{\bm{e}}_ZD_X^T=\bm{s}_X$ and
$\bar{\bm{e}}_XD_Z^T=\bm{s}_Z$.

Retaining the original variables and enforcing
Eq.~(\ref{eq:gari_effective}) gives the augmented matrix
\begin{equation}
\bar{D}_{XYZ}=\begin{blockarray}{cccccc}
\bm{e}_Z & \bm{e}_X & \bm{e}_Y & \bar{\bm{e}}_Z & \bar{\bm{e}}_X\\
\begin{block}{(c|c|c|c|c)c}
  \bm{0}  &  \bm{0}  & \bm{0}  &    D_X    &  \bm{0}  & \bm{s}_X  \\
  \bm{0}  &  \bm{0}  & \bm{0}  &   \bm{0}  &   D_Z    & \bm{s}_Z  \\
     I    &  \bm{0}  &    U    &    I      &  \bm{0}  & \bm{0} \\ 
  \bm{0}  &    I     &    V    &   \bm{0}  &   I      & \bm{0} \\
\end{block}
\end{blockarray}.
\label{eq_gari}
\end{equation}
The upper row blocks enforce the measured detector outcomes.  The
lower blocks are virtual checks with zero syndromes, which enforce the
auxiliary-variable definition in Eq.~(\ref{eq_gari}). The resulting
graph represents the same decoding problem but removes the four-cycles
involving the mixed-error variables. BP can consequently behave
differently on the transformed graph, even though the underlying error
constraints are equivalent. GARI thus provides a graph representation
on which BP-based decoding, including BP-SF, can operate.

\subsection{Sliding-Window Decoding}
\label{sec:window_decoding}

Repeated syndrome extraction produces a decoding graph that grows
along the time direction. Monolithic decoding processes the entire
graph at once, whereas sliding-window decoding processes overlapping
temporal regions and finalizes older corrections incrementally. This
approach has also been studied for qLDPC codes~\cite{huang2024sliding,
  gong2024toward}.

\begin{figure}[t]
    \centering
    \includegraphics[width=\singlefigwidth]{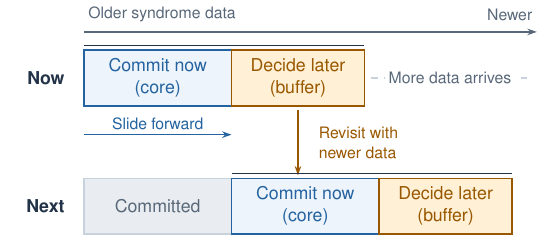}
    \caption{Schematic workflow for sliding-window decoding. Decode
      the core and buffer together, but finalize only the older core
      estimates. As new syndrome data arrive, slide the window forward
      and revisit the buffer with more context.}
    \label{fig:window_procedure}
\end{figure}

A window contains $W=C+B$ time blocks into an older \emph{core} of
$C$ blocks and a newer \emph{buffer} of $B$ blocks, as illustrated in
Fig.~\ref{fig:window_procedure}. The decoder estimates errors in both
regions but commits only the core estimates, meaning that these
decisions are finalized. The buffer provides later detector outcomes
that help resolve errors near the core boundary.  Its estimates remain
tentative and are reconsidered when the window advances. In this way,
each decoding step uses more temporal context than the region for
which it finalizes decisions.

Before decoding the next window, the contribution of the committed
correction must be removed from the syndrome. The resulting
\emph{residual syndrome} contains the detector outcomes still
unexplained by the committed estimates.  Using the complete syndrome
record $\bm{s}$ for notation, let $\bm{r}^{(j)}$ denote this residual
before window $j$, with $\bm{r}^{(0)}=\bm{s}$.  If
$\hat{\bm{e}}_j^{\mathrm{com}}$ contains the newly committed
estimates, extended by zeros outside the core, then
\begin{equation}
\label{eq:window_residual}
    \bm{r}^{(j+1)}
    = \bm{r}^{(j)}
    + \hat{\bm{e}}_j^{\mathrm{com}}H^T.
\end{equation}
Here, $H$ is the detector--error mechanism matrix.  The update
includes the committed correction's effects on buffer checks, so that
subsequent windows account for decisions already made. A streaming
decoder evaluates this relation only for detector outcomes that have
arrived. The window advances by $C$ blocks after each core is
committed.  At the final extraction boundary, the remaining tail is
decoded and committed using the available boundary information.
Successive windows are processed sequentially in this scheme, while
BP-SF trials can run in parallel within each window.

\section{BP Behavior Analysis: A Case Study}

We use the $\llbracket144,12,12\rrbracket$ ``gross'' code
from~\cite{Bravyi_2024} as a concrete example for studying BP on qLDPC
systems. The purpose of the case study is to expose patterns that can
guide decoder design. We use the min-sum update in Eq.~(\ref{c2v})
with the iteration-dependent normalization factor $\alpha=1-2^{-i}$, where $i$ denotes the
current iteration. This setting is used consistently in the analysis
that follows.

\subsection{Number of Iterations}

\begin{figure}[t]
    \centering
    \includegraphics[width=\singlefigwidth]{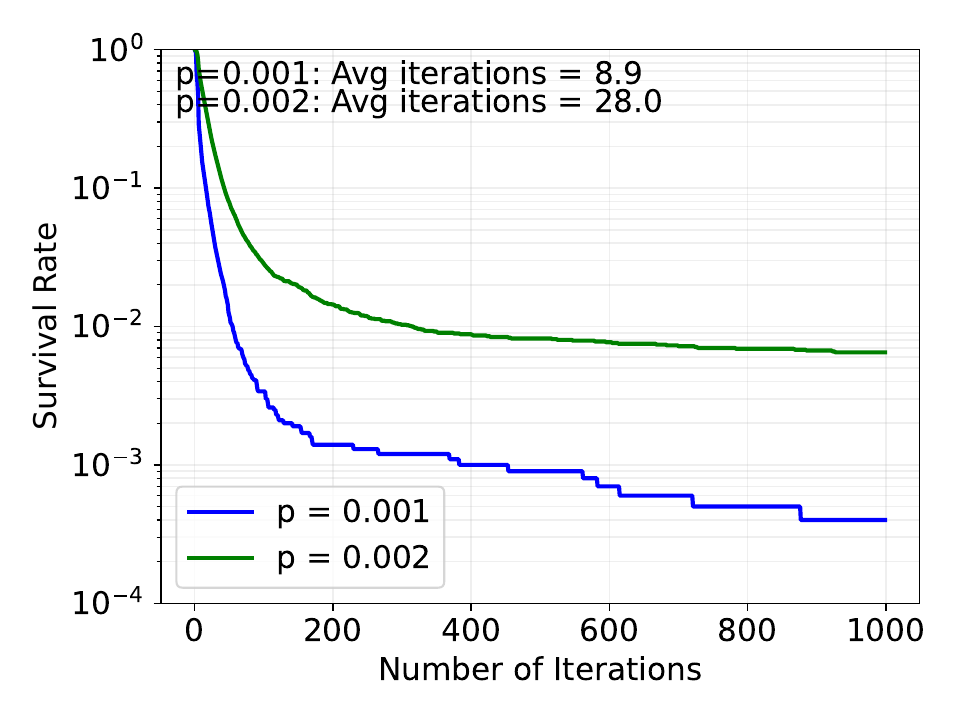}
    \caption{BP convergence survival on the
      $\llbracket144,12,12\rrbracket$ code under circuit-level noise
      with the fraction of samples not yet converged by each
      iteration. We use 10,000 samples per physical error rate and a
      maximum of 1,000 iterations.}
    \label{idist}
\end{figure}

Fig.~\ref{idist} measures how quickly BP resolves syndromes under
circuit-level noise. We simulate 10{,}000 samples at each of $p=0.001$
and $p=0.002$, representative physical error rates below threshold,
and record the iterations required by every sample. We plot the
survival rate, defined as the fraction of samples that have not
converged by iteration $i$, equal to one minus the cumulative
convergence rate. Survival here refers to BP non-convergence and the
tail at the iteration limit counts samples that exhaust the budget
without converging.

The distribution in Fig.~\ref{idist} shows that most syndromes
converge after only a few iterations. At $p=0.001$, the mean is 8.9
iterations even though the decoder allows up to 1,000. Increasing the
physical error rate to $p=0.002$ lengthens the tail but does not
substantially change the typical case. Samples that survive the early
iterations rarely become easier merely by increasing the iteration
limit. This motivates changing the decoder input instead of repeatedly
extending one BP trajectory. If the perturbed inputs offer partially
independent opportunities for success, parallel trials can lower the
logical error rate with a latency determined by the slowest active
stage and the available hardware.

\subsection{Oscillation}

\begin{figure}[t]
    \centering
    \includegraphics[width=\singlefigwidth]{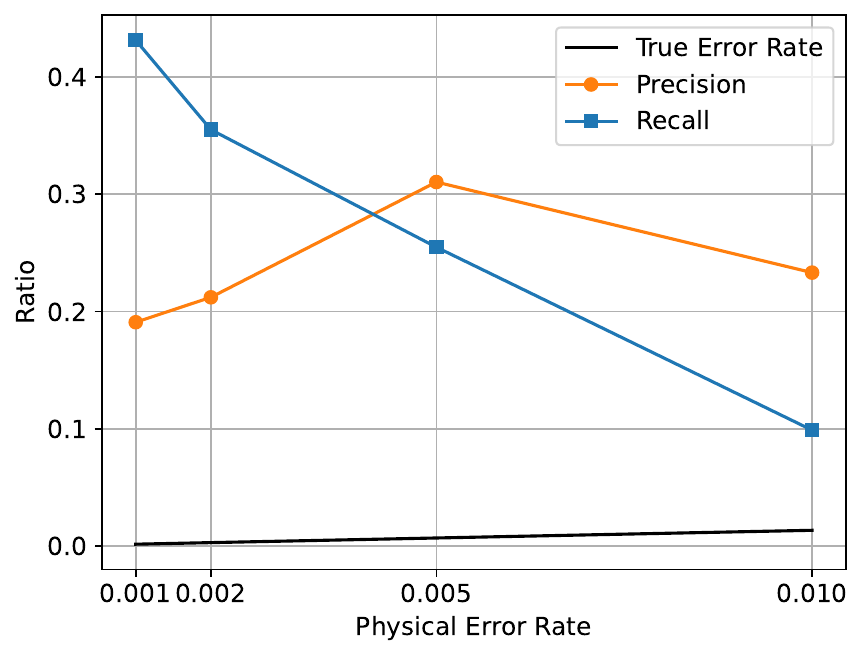}
    \caption{Precision and recall probabilities of candidate bit
      selection on the $\llbracket144,12,12\rrbracket$ code. We
      evaluate the correlation between candidate bits and actual error
      locations by identifying the top 50 most frequently flipped bits
      among approximately 8,700 error mechanisms (fewer than 50 are
      selected if some of the top-ranked bits are never flipped). The
      decoder is run with a maximum of 50 iterations, and statistics
      are collected over 1,000 decoding failures. This analysis
      reveals how well bit-level oscillation can serve as a heuristic
      for error localization.}
    \label{osc}
\end{figure}

Prior work~\cite{raveendran2021trapping, 10409156} links BP failures
to trapping sets, degeneracy, and the resulting ambiguity of iterative
messages. Chytas et al.~\cite{chytas2025iteration} analyze min-sum
message oscillations associated with symmetric stabilizers under an
isolation assumption and exploit previous-iteration messages to modify
the update rule. Here, we instead examine hard-decision oscillations
as a signal for selecting post-processing trials. We track how often
output bits alternate between zero and one to determine whether they
identify the locations most relevant to a decoding failure.  For each
failed run, we count how often every bit flips between consecutive
hard decisions and select the highest-count bits as the candidate set
$\Phi$. We denote the support of the true error by
$\operatorname{supp}(\bm{e})$ and measure the overlap of $\Phi$ with
that support using precision and recall:
\begin{align}
  \text{Precision} &= \frac{|\operatorname{supp}(\bm{e}) \cap \Phi|}{|\Phi|},  \\
  \text{Recall} &= \frac{|\operatorname{supp}(\bm{e}) \cap \Phi|}{|\operatorname{supp}(\bm{e})|}.
\end{align}
Fig.~\ref{osc} reports precision and recall when min-sum BP fails to
decode a syndrome. Even when the final estimate is incorrect, the
oscillation pattern identifies a useful subset of the actual error
locations, especially at low physical error rates. The precision of
the oscillation set is much larger than the raw physical error rate,
making $\Phi$ a more informative target than random guessing. As the
physical error rate rises, recall falls because the number of errors
grows while the candidate-set size remains fixed.

To test whether oscillations improve decoding rather than merely
correlate with errors, we compare two ensembles that reverse one prior
LLR per additional trial, equivalently $p'_{v_i}\leftarrow1-p_{v_i}$,
which can also be proved to be equivalent to flipping the supporting
syndrome bits. Candidate positions are selected either from the most
oscillating bits or at random. This comparison tests where to perturb
while keeping the perturbation rule fixed.

\begin{figure}
    \centering
    \includegraphics[width=\singlefigwidth]{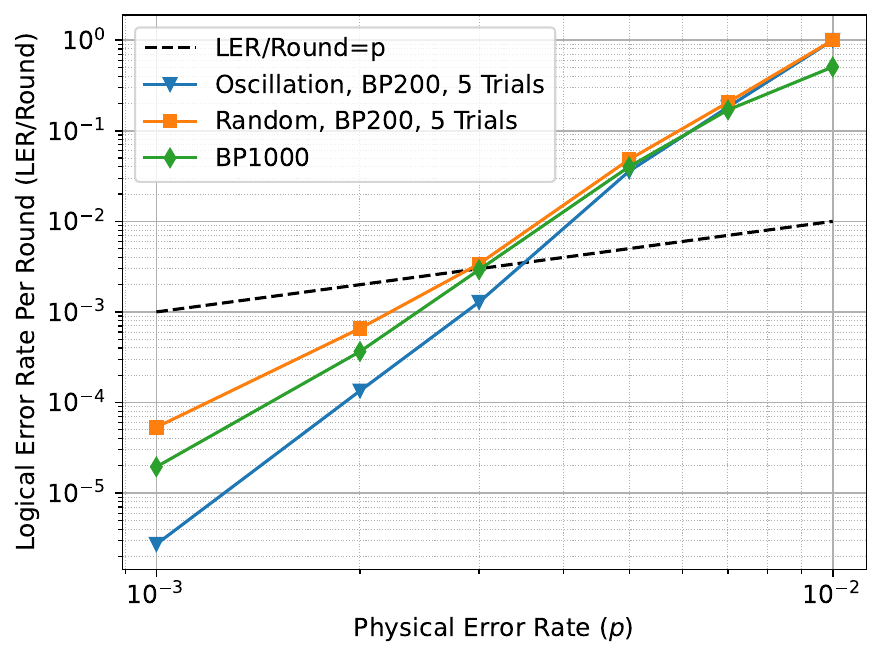}
    \caption{Effect of candidate selection on the
      $\llbracket144,12,12\rrbracket$ code under circuit-level
      noise. ``5 Trials'' denotes one initial BP attempt and up to
      four additional trials upon failure, with at most 200 iterations
      per attempt. Each additional trial reverses one prior
      LLR. Candidate positions are the four most oscillating bits or
      are selected at random. BP1000 uses a single attempt with at
      most 1,000 iterations.}
    \label{fig:rndvsosc}
\end{figure}

Fig.~\ref{fig:rndvsosc} shows, for different physical errors rates $p$
(x-axis) that oscillation-guided selection (blue) lowers the logical
error rate (LER) per round (y-axis) relative to random selection (red)
and BP with 1,000 iterations (green) with the same configured trial
budget, especially at low $p$. Thus, candidate selection matters
beyond simply adding decoding attempts. This result motivates using
oscillation-guided candidate selection to construct the independent
trials described next.

\section{Oscillation-Guided Syndrome-Flipping Decoding}

The candidate-selection result motivates the BP-SF procedure
introduced in our prior work~\cite{wang2026bpsf}, which we review here
as the basis for the GPU and windowed extensions.

As illustrated in Fig.~\ref{scheme}, our decoder returns the first
syndrome-valid result after reversing its trial perturbation,
restoring consistency with the original
syndrome. Algorithm~\ref{alg:mybp} gives the complete procedure. More
specifically, during the initial BP attempt, we record the oscillation
score
\begin{equation}
\label{eq:oscillation_score}
    \omega_i =
    \sum_{t=1}^{I_0}
    \left(\hat e_i^{(t)} \oplus \hat e_i^{(t-1)}\right),
\end{equation}
where $I_0$ is the initial iteration limit, $\hat e_i^{(0)}$ is the
hard decision from the channel LLR, and the sum counts changes using
ordinary integer arithmetic.  If the attempt does not converge,
$\operatorname{TopK}$ selects up to $K$ eligible variables with the
highest positive scores.  For each retry, $\operatorname{Sample}$
generates a binary trial vector supported on $\Phi$, with a weight
between one and $\min(w_{\max},|\Phi|)$. The allocation of retries
among flip weights is specified by the experimental configuration. The
final XOR reverses the trial perturbation: Whenever
$\hat{\bm e}_r H^T=\bm s_r$, the returned estimate satisfies
$(\hat{\bm e}_r\oplus\bm t_r)H^T=\bm s$.  Thus, each trial explores a
perturbed decoding problem while its accepted output remains
consistent with the original syndrome. All decoder calls use the same
matrix $H$ and channel LLRs.  $\operatorname{InitialBP}$ returns a
convergence flag, an error estimate, and the oscillation scores.
$\operatorname{TrialDecode}$ runs the configured BP or DMem-BP decoder
for at most $I_{\mathrm{tr}}$ iterations, including the configured
memory initialization from the initial attempt.  A global return
selects the first converged trial and stops the remaining trials.

The motivation behind this first-convergence stopping is as follows:
In the data-qubit Pauli picture, two estimates with the same syndrome
differ by either a stabilizer or a nontrivial logical operator, and in
the latter case, correction is off at least at $d$
locations. Consequently, a syndrome-consistent estimate whose
difference from the most likely error has weight below $d$ implements
the same logical correction, even if its individual likelihood is
lower. A similar first-convergence rule is used by the GARI
ensemble~\cite{maan2025decoding}. In addition, the results
in~\cite{muller2025improved} also support that the return of waiting
for more completes is less significant.

We note that targeted trial generation is related to many
ensemble-based decoding algorithms, e.g., Chase
decoding~\cite{1054746}, which constructs test patterns from channel
reliabilities for repeated hard-decision
decoding. AutDEC~\cite{koutsioumpas2025automorphism} also transforms
syndromes, but uses code automorphisms to construct its BP
ensemble. Multiple-bases BP list decoding~\cite{rabeti2025list}
instead varies the parity-check representation and applies a decision
rule to the resulting candidate list. In general, ensembles are a
standard way to improve decoders by generating several modified inputs
and selecting a result using a likelihood or weight
criterion~\cite{1054746, 4036366, 10.1109/TCOMM.2009.07.070621,
  WANG2023102194}. Independent trials can execute concurrently, so
their cost can be exchanged for latency when hardware resources are
available.

\begin{algorithm}[t]
\caption{Oscillation-guided BP-SF}
\label{alg:mybp}
\SetKwInOut{Input}{Input}
\SetKwFor{ForPar}{parallel for}{}{end}
\SetKw{ReturnAll}{return globally}

\Input{Syndrome $\bm{s}$; candidate limit $K$;
maximum flip weight $w_{\max}$; retry budget $N_{\mathrm{tr}}$;
iteration limits $I_0,I_{\mathrm{tr}}$}
\KwResult{Syndrome-consistent estimate or decoding failure}

$(\mathrm{converged},\hat{\bm e},\bm\omega)
    \gets \operatorname{InitialBP}(\bm s,I_0)$\;
\If{$\mathrm{converged}$}{
    \Return $\hat{\bm e}$\;
}

$\Phi \gets \operatorname{TopK}(\bm\omega,K)$\;
\If{$\Phi=\emptyset$}{
    \Return decoding failure\;
}

\ForPar{$r=1,\ldots,N_{\mathrm{tr}}$}{
    $\bm t_r \gets \operatorname{Sample}(\Phi,w_{\max})$\;
    $\bm s_r \gets \bm s \oplus \bm t_r H^T$\;
    $(\mathrm{converged}_r,\hat{\bm e}_r)
        \gets \operatorname{TrialDecode}(\bm s_r,I_{\mathrm{tr}})$\;
    \If{$\mathrm{converged}_r$}{
        \ReturnAll $\hat{\bm e}_r \oplus \bm t_r$\;
    }
}
\Return decoding failure\;
\end{algorithm}

\section{Implementations}
\label{sec:gpu_impl}
\begingroup
GPU implementations of BP exploit parallel message updates, sparse
graph storage, and concurrent
decoding~\cite{dai2022multigbps,ferraz2025gpu}.  We use these
established techniques to implement the initial BP attempt and
speculative trials of Algorithm~\ref{alg:mybp} in CUDA.  The backend
uses single-precision floating-point messages and exposes a Python
interface for configuring the decoder and submitting syndromes.

The parity-check matrix is stored on the device in compressed sparse
row form, together with a variable-oriented adjacency index and an
edge map for accessing the same message arrays from either direction.
The graph and channel LLRs remain on the device, and message buffers
are allocated once and reused across decoding requests. During
flooding iterations, small groups of threads cooperate on each
check-node update, using warp shuffle operations to reduce message
magnitudes and signs.  Variable-node updates distribute the output-LLR
and outgoing-message calculations across threads. The initial BP stage
executes its iteration loop within a cooperative kernel, using
grid-wide barriers between dependent update phases and tracking
hard-decision changes on the device.

If the initial attempt fails, the host launches device-side selection
of the $|\Phi|$ most oscillating bits, followed by the speculative
stage.  Each trial is assigned to one CUDA block, which chooses its
flip positions, constructs the perturbed syndrome using column
adjacency lists, and runs BP with its own message state. All trials
share the read-only graph and channel LLRs. A trial's syndrome is kept
in shared memory when it fits the configured shared-memory budget,
with a global-memory buffer used otherwise. Block-wide barriers
enforce the update order within each trial. The BP iteration loop
requires no host intervention.

A shared device-wide success flag implements BP-SF's first-success
selection rule. A block whose estimate passes the syndrome check
claims the result using an atomic compare-and-swap operation, writes
the correction, and reverses its trial flips. Other blocks check the
flag at iteration boundaries and stop when a result is available.  The
host retrieves the result after the kernel completes.

For configurations using DMem-BP, each trial additionally maintains
the previous output LLRs and fixed node-specific memory strengths
$\gamma_{v_i}$, applying Eq.~(\ref{eq:dmem_bias}) during variable-node
updates. The previous output LLRs can be initialized from the channel
LLRs or from the failed initial BP attempt, where edge messages
restart from the channel LLRs in both cases. The trials then evolve
independently.  This implementation supports evaluating syndrome
flipping with either ordinary BP or DMem-BP as the component
decoder. The source code is available at
\href{https://github.com}{GitHub(TBD)}.
\endgroup

\section{Simulation Results}
\begingroup
We evaluate BP-SF from individual decoding calls to streaming quantum
memory. The experiments first establish its accuracy--latency
trade-off, then examine how windowing, graph representation, and
message scheduling shape the time at which corrections become
available.

\subsection{Experimental Setup}
\label{sec:simulation_setup}

We generate syndrome-extraction circuits with
Stim~\cite{gidney2021stim} and apply circuit-level noise, including
faults in gates and measurements.  Figures~\ref{fig:complexity_growth}
and~\ref{fig:window_scaling} use fixed-basis memory
experiments. Figure~\ref{fig:arbitrary_state} evaluates both
components of the logical Pauli frame, using ideal initial and final
$X/Z$ stabilizer extraction without destructive data readout. These
two protocols have different boundary and success conditions.
Comparisons of decoder accuracy are made within each protocol. The
baseline comparison uses $R=d$ noisy extraction rounds, and the
long-memory studies extend $R$ to 512.

For a memory experiment of $R$ noisy rounds, we report the effective
logical error rate per round as
\begin{equation}
    \label{eq:ler_per_round}
    \mathrm{LER}_{\mathrm{round}}=1-(1-P_{\mathrm{fail}})^{1/R},
\end{equation}
where $P_{\mathrm{fail}}$ counts both decoder non-convergence and converged
corrections with incorrect logical observables. The normalization is per
noisy round, rather than per logical qubit. LER error bars are nominal
95\% Wilson intervals transformed to this scale.

All GPU timing results use an NVIDIA GeForce RTX 5060 Ti. The survival
rate estimates the probability of a shot that is still under decoding
at time $t$, defined as
$\Pr(T_{\mathrm{call}}>t)$. Figure~\ref{fig:complexity_growth}(c)
depicts calls on samples for which every tested method's native
initial BP stage fails within 80 iterations, and
Fig.~\ref{fig:arbitrary_state}(b) reports unfiltered samples. For
windowed operation, we distinguish decoder compute time from
correction-release age, which also includes waiting for core/buffer
data and queuing.  Accuracy and timing use separate sample
sets. Timing boundaries and sampling procedures are detailed in
Appendix~\ref{app:simulation_details}.

\subsection{Accuracy-Latency Trade off on Monolithic Decoding}
\label{sec:ler_latency}
Figure~\ref{fig:complexity_growth} compares BP-SF with plain BP,
BP-OSD, and Relay-BP~\cite{muller2025improved}. BP-SF combines an
ordinary BP attempt with 100 posterior-initialized DMem-BP
retries. The accuracy comparison uses a single Relay-BP chain of 300
memory-parameter sets, BP1000 allows 1,000 BP iterations, and
BP1000-OSD10 adds an order-10 OSD.

\begin{figure*}[tp]
    \centering
    \includegraphics[width=\linewidth]{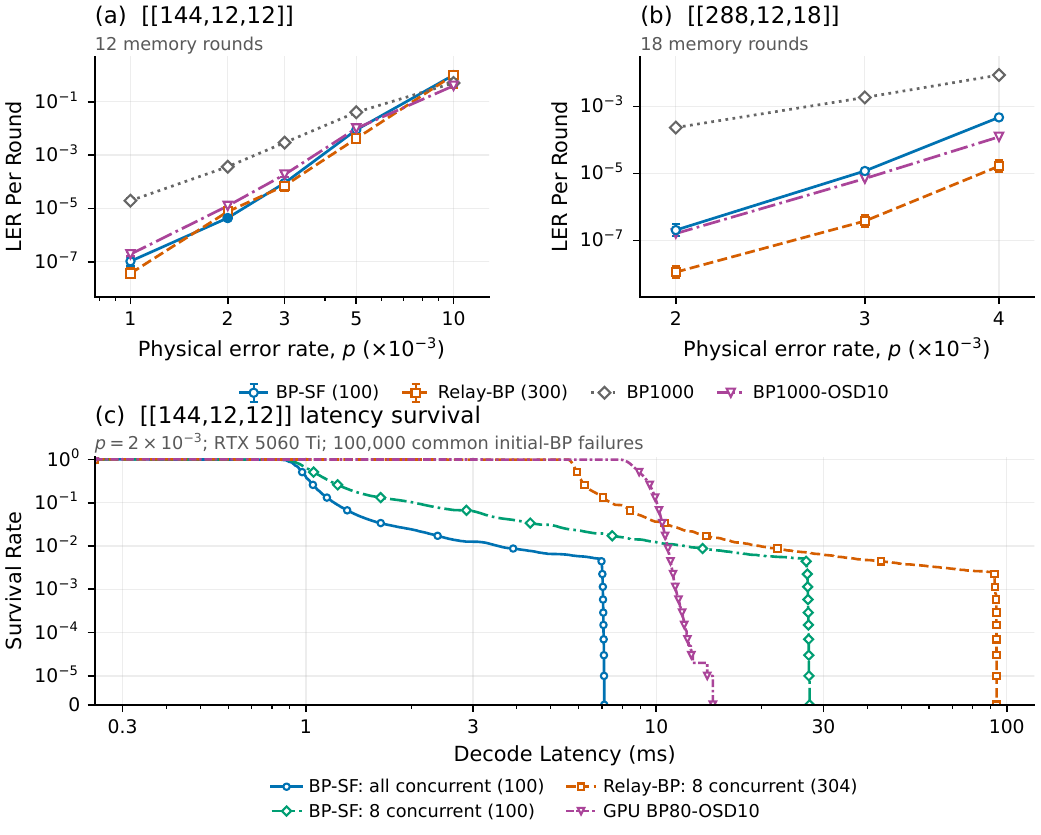}
    \caption{Accuracy and call latency of BP-SF and baseline
      decoders; parentheses give retry budgets. (a,b) Effective LER per round
      for the $\llbracket144,12,12\rrbracket$ and
      $\llbracket288,12,18\rrbracket$ codes, using 12 and 18 memory
      rounds. (c) Latency survival on 100,000 matched 144-code samples
      at $p=0.002$, conditioned on every tested method's native initial
      BP stage failing within 80 iterations. Survival is the fraction
      of complete scalar calls still running after time $t$, including
      the initial BP stage and all return outcomes.
    }
    \label{fig:complexity_growth}
\end{figure*}

On the $\llbracket144,12,12\rrbracket$ code, BP-SF and Relay-BP
achieve LERs within approximately a factor of three over
$p=0.001$--$0.005$.  At $p=0.002$, BP-SF reaches $4.38\times10^{-6}$
per round, an $83\times$ reduction relative to BP1000 and a
$2.83\times$ reduction relative to BP1000-OSD10. On the
$\llbracket288,12,18\rrbracket$ code, Relay-BP is more accurate:
BP-SF's LER is $18$--$31\times$ higher at the three tested error rates
and also exceeds the BP1000-OSD10 reference.  The following latency
and system-level studies focus on the 144 code, where BP-SF provides a
competitive accuracy operating point.

Parallel retries shorten the long-running calls in
Fig.~\ref{fig:complexity_growth}(c). Every selected sample exhausts
each decoder's initial BP budget, so the plotted population contains
no initial BP early returns. We see a sharp descent in BP-OSD
because of comparatively regular OSD workload at fixed matrix size and
OSD order. We note that all 100,000 BP80-OSD10 calls return syndrome-consistent corrections, but 2,719 have the wrong logical effect. Thus, its final descent to
zero indicates completion, not logical correctness. In contrast, BP-based results usually either returns a correction in the correct logical coset, or exhausts all iterations and aborts.

On this common initial-failure population, all-concurrent BP-SF has
a mean call latency of $1.081$ ms and a 99.9th-percentile latency of
$7.039$ ms, compared with $6.796$ ms and $92.556$ ms for eight-chain
Relay-BP. These are reductions of $6.28\times$ and $13.15\times$,
respectively. GPU BP80-OSD10, including its 80-iteration initial BP
stage, requires $9.059$ ms and $11.347$ ms. Relative to BP80-OSD10,
all-concurrent BP-SF reduces mean and p99.9 latency by $8.38\times$
and $1.61\times$, respectively. We choose eight lanes because the
CUDA-Q~\cite{cudaq} Relay-BP implementation used here supports at
most eight parallel lanes. Restricting BP-SF to eight concurrent
trials raises its mean and p99.9 latency to $1.556$ ms and
$27.051$ ms, respectively, still below the eight-lane Relay-BP
timings. Its p99.9 latency is nevertheless higher than BP80-OSD10's
$11.347$ ms. Additional BP-SF concurrency therefore benefits the
latency tail more strongly than the mean. The next experiment asks
whether this short-memory operating point can be extended to much
longer syndrome histories.

\subsection{Temporal Scaling Using Window Decoding}
\label{sec:window_scaling}
A longer memory produces a larger decoding graph, even when the code
remain fixed.  We compare monolithic BP-SF with the sliding-window
procedure of Section~\ref{sec:window_decoding} on the 144 code at
$p=0.002$. Both use up to 80 initial BP iterations and 60 iterations
per DMem-BP retry. The 100-retry limit applies once per memory
experiment for monolithic decoding and separately to each window for
windowed decoding. Within a window, candidate selection and local
convergence checks use the core. Appendix~\ref{app:simulation_details}
gives the remaining decoder settings.

We first choose the core and buffer lengths using the 128-round sweep
in Table~\ref{tab:window_geometry}. Increasing the buffer from one to
three rounds lowers the measured LER at every tested core length;
extending it to six gives little further change. At $C=7$, for
example, LER falls from $2.95\times10^{-5}$ to $1.19\times10^{-5}$
when $B$ increases from one to three, while $B=6$ gives
$1.20\times10^{-5}$ and adds roughly 9 ms to p99 correction-release
age. We therefore use a three-round buffer.

\begin{table}[t]
    \centering
    \caption{Core ($C$) / buffer ($B$) sweep for windowed BP-SF on the
      $\llbracket144,12,12\rrbracket$ code at $p=0.002$ and 128
      rounds.  Release age includes core/buffer waiting, queuing, and
      decoding.  Bold entries mark the geometry used in
      Fig.~\ref{fig:window_scaling}.}
    \label{tab:window_geometry}
    \begingroup
    \small
    \setlength{\tabcolsep}{3pt}
    \renewcommand{\arraystretch}{1.1}
    \begin{tabular*}{\columnwidth}{@{\extracolsep{\fill}}ccrrr@{}}
        \hline
        $C$ & $B$ & LER/round & Mean compute & p99 age \\
        & & ($10^{-5}$) & ($\mu$s/round) & (ms) \\
        \hline
        4 & 1 & 5.26 & 45.94 & 12.52 \\
        4 & 3 & 1.59 & 42.15 & 18.32 \\
        4 & 6 & 1.53 & 42.63 & 27.21 \\
        7 & 1 & 2.95 & 38.71 & 21.38 \\
        \textbf{7} & \textbf{3} & \textbf{1.19} & \textbf{35.42} & \textbf{27.21} \\
        7 & 6 & 1.20 & 38.27 & 36.17 \\
        10 & 1 & 2.27 & 34.57 & 30.24 \\
        10 & 3 & 0.94 & 34.51 & 36.17 \\
        10 & 6 & 0.95 & 37.57 & 45.12 \\
        \hline
    \end{tabular*}
    \endgroup
\end{table}

With $B=3$, increasing the core from four to seven rounds reduces
compute time from $42.15$ to $35.42~\mu$s/round. A ten-round core
saves only another 2.6\%, while raising p99 release age from $27.21$
to $36.17$ ms.  Although a ten-round core gives a lower LER point
estimate, a seven-round core retains most of its compute efficiency
and releases corrections earlier. We use $C=7,B=3$ for the length
sweep to balance these costs.

\begin{figure*}[t]
    \centering
    \includegraphics[width=\textwidth]{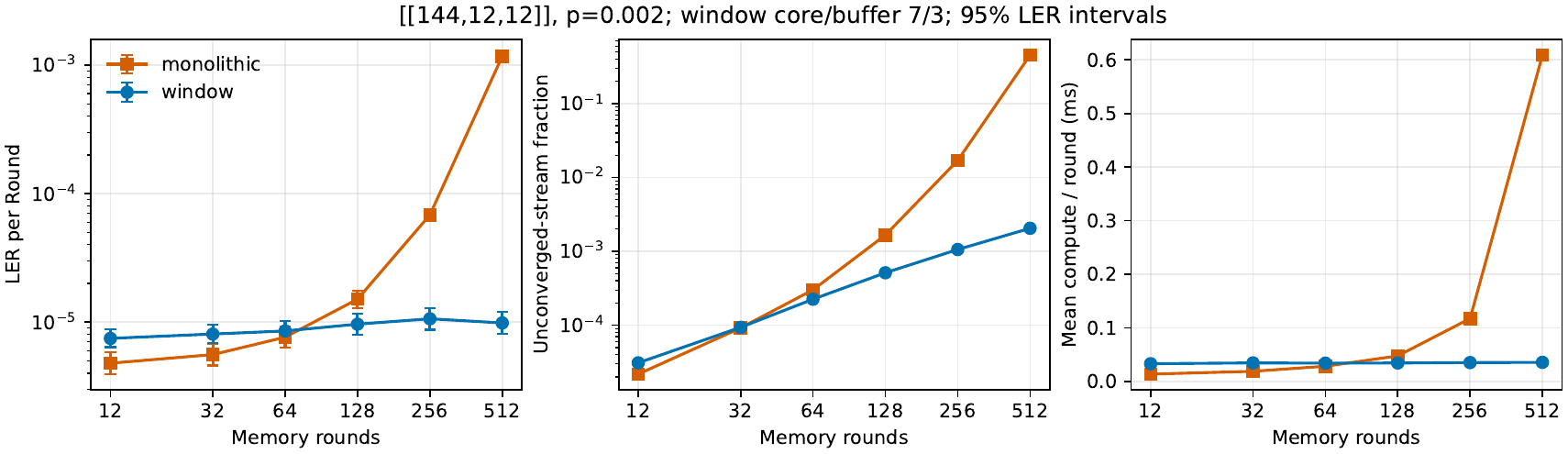}
    \caption{Temporal scaling of BP-SF on the
      $\llbracket144,12,12\rrbracket$ code at $p=0.002$, comparing
      monolithic decoding with $7/3$ window decoding. Left: effective
      LER per round. Middle: fraction of complete memory experiments
      that fail to converge. Right: mean decoder compute time per
      round, timed for each shot. Only compute time is recorded and
      syndrome lookahead time are excluded.}
    \label{fig:window_scaling}
\end{figure*}

With this geometry, windowed BP-SF maintains an effective LER near
$10^{-5}$ through 512 rounds (see
Fig.~\ref{fig:window_scaling}). Monolithic decoding is more accurate
for short memories, but its LER rises sharply on longer
graphs. Non-convergence accounts for 9,245 of its 9,246 failed memory
experiments at 512 rounds, making it the dominant source of error.

The mean windowed compute time remains between $0.033$ and $0.036$ ms
per round across 12--512 rounds. Monolithic decoding is faster through
64 rounds, but windowing becomes faster between 64 and 128 rounds and
reaches a $17.1\times$ speedup at 512 rounds. This near-constant
per-round cost makes windowed BP-SF a useful basis for long-memory
decoding.  Section~\ref{sec:arbitrary_state} next examines full-frame
recovery, connecting graph and schedule choices to decoding accuracy,
call latency, and correction release during streaming memory.

\subsection{Full-Frame Error Correction}
\label{sec:arbitrary_state}

Preserving an arbitrary logical state requires recovering both
components of its Pauli frame. A residual logical $Z$, for example,
leaves $\ket{0}_L$ unchanged but can corrupt a superposition. We
therefore extend the evaluation from fixed-basis memory to full-frame
recovery on the 144 code, checking all 24 logical $X/Z$ labels on
every sample. The full-frame protocol uses ideal $X/Z$ extraction at
both memory boundaries and no destructive data readout. Its
construction and scoring rule are given in
Appendix~\ref{app:simulation_details}.

We consider three ways to recover the full Pauli frame within
BP-SF. The first decodes the two sectors sequentially using $D_Z$ and
$D_X$. After estimating the effective $X$ errors with $D_Z$, we update
the $Z$-error priors by Bayesian conditioning on the accepted
estimate, and then decode with $D_X$.  This approach exploits
cross-sector correlations while retaining the smaller single-sector
graphs, following the principle of correlation-aware CSS
decoding~\cite{delfosse2014correlations}.  The second uses the joint
matrix $D_{XYZ}$, which explicitly includes mixed-error mechanisms and
allows information to pass between the two sectors throughout
decoding.  The third applies GARI to the joint model, introducing
auxiliary variables and consistency checks to reorganize these
connections and remove the targeted short
cycles~\cite{maan2025decoding}.  All three approaches use BP-SF, but
they differ in how the decoding problem represents and exchanges
correlated error information.

These choices affect both inference and computational cost.  With
standard message aggregation, a complete BP iteration requires
$O(|\mathcal E|)$ arithmetic work, where
$|\mathcal E|=\operatorname{nnz}(D)$ is the number of edges in the
Tanner graph of the decoding matrix $D$.  Joint decoding retains more
error information in a single graph, but changes the number and
connectivity of messages that must be updated. GARI further changes
the decoding graph by enlarging it but making the graph sparser.  The
total decoding cost also depends on the iterations and retries needed
to converge, while GPU latency depends on the message schedule and
available parallelism. For example, in our twelve-round full-frame
experiment on the $\llbracket144,12,12\rrbracket$ code, $D_X$ and
$D_Z$ each contain $30{,}672$ edges, giving $61{,}344$ edges across
the two separate graphs. The joint matrix $D_{XYZ}$ contains
$419{,}904$ edges, whereas the GARI matrix contains $203{,}904$.
Thus, joint and GARI decoding involve approximately $6.85\times$ and
$3.32\times$ as many edges, respectively, as the two separate graphs
combined.  Despite having more rows and columns, GARI reduces the edge
count by $51.4\%$ relative to the joint representation.

We therefore compare these representations on the
$\llbracket144,12,12\rrbracket$ code at $p=0.002$ and $R=12$, asking
whether their gains in full-frame accuracy translate into favorable
latency once graph cost, convergence, and retry budgets are
considered, especially when these representations causes different BP
complexity, which could benefit the LER on small graphs by sending
more retries within the same latency budget.

For the $D_{XYZ}$ and GARI graphs, we compare flooding with conflict-free
physical-check layers of at most 128 checks, labeled ``ind.128.'' The
layers are fixed, their order is shuffled each sweep, and GARI's two
virtual layers follow in fixed $U,V$ order. More information on
layered decoding for GARI can be found
in~\cite{bascones2026gari,maan2025decoding}. In the comparisons below,
$D_{XYZ}$ without a schedule suffix denotes $D_{XYZ}$ with flooding
updates, matching the figure labels; $D_{XYZ}$ ind.128 denotes the
layered configuration.

\begin{figure*}[!t]
    \centering
    \includegraphics[width=\textwidth]{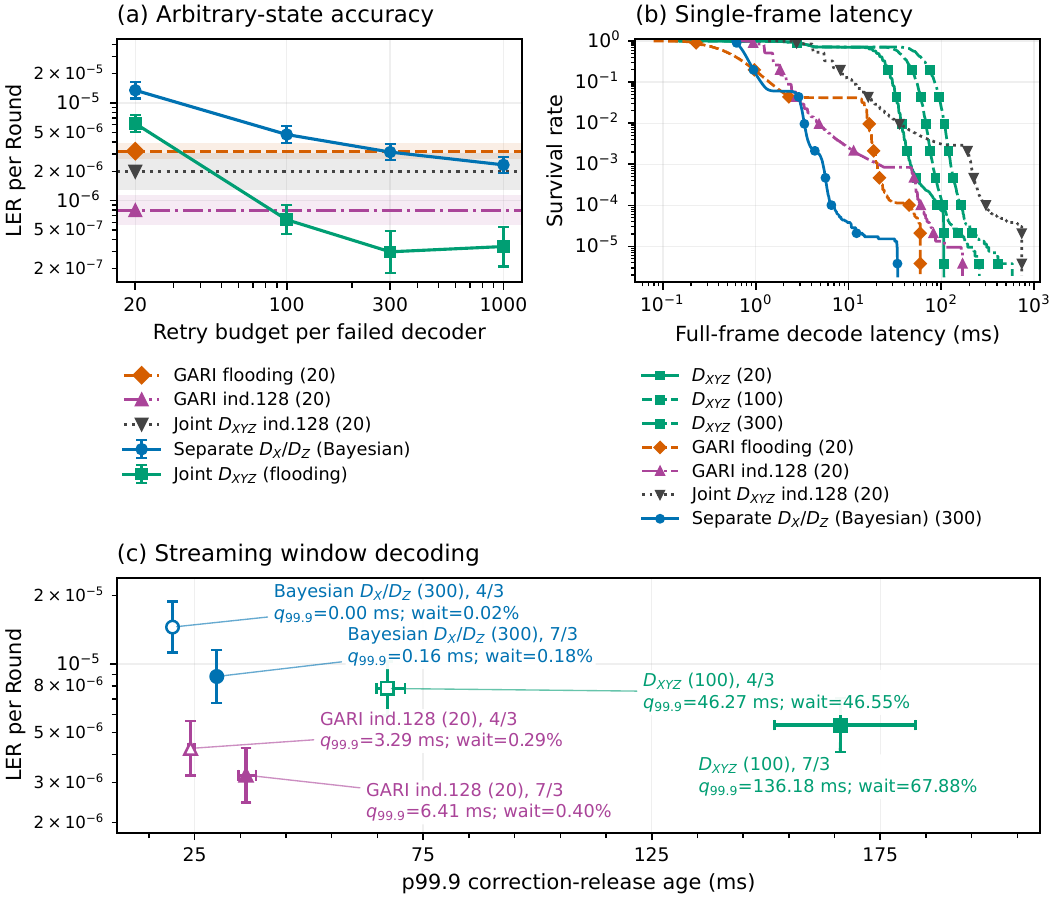}
    \caption{Full-frame recovery with BP-SF on the
      $\llbracket144,12,12\rrbracket$ code at $p=0.002$. Parentheses
      give retry budgets. (a) Twelve-round memory accuracy for
      separate, joint, and GARI graph representations.  (b)
      Call-latency survival on 524,288 matched, unfiltered samples per
      configuration. ind.128 denotes a layer-size cap of 128 checks.
      (c) 512-round streaming memory with 2.97-ms syndrome arrivals.
      Filled/open markers denote $7/3$ and $4/3$ core/buffer
      geometries. $q_{99.9}$ is p99.9 computational queue delay,
      excluding lookahead and own processing; ``wait'' is the queued
      arrival fraction. Bars and bands show nominal 95\% LER intervals and 95\%
      release-age intervals.  Timing in (b,c) uses RTX 5060 Ti GPU.
    }
    \label{fig:arbitrary_state}
\end{figure*}

Joint decoding makes better use of the retry budget in this full-frame
experiment as shown Fig.~\ref{fig:arbitrary_state}(a). Separate
decoding reaches $2.32\times10^{-6}$ LER/round with 1,000 retries per
sector, whereas $D_{XYZ}$ reaches $6.36\times10^{-7}$ with 100
retries. Increasing the number of separate-sector trials therefore
does not close the measured accuracy gap to joint decoding.

The schedule also matters. With 20 retries, changing GARI from
flooding to ind.128 lowers its LER from $3.21\times10^{-6}$ to
$7.95\times10^{-7}$. GARI ind.128 with 20 retries and $D_{XYZ}$
with 100 retries have similar measured LERs and overlapping 95\%
intervals, making them useful operating points for the latency
comparison. $D_{XYZ}$ can reach a lower LER of
$2.98\times10^{-7}$ with 300 retries, at a larger trial budget.

Figure~\ref{fig:arbitrary_state}(b) shows the timing consequence of
these choices on unfiltered samples. GARI ind.128 with 20 retries
takes $1.624$ ms on average, compared with $31.304$ ms for
$D_{XYZ}$ with 100 retries, a $19.3\times$ reduction. Their p99.9
latencies are $20.983$ and $80.843$ ms, respectively, a $3.85\times$
reduction.  The graph and schedule therefore offer a substantially
faster BP-SF operating point at similar measured accuracy. We carry
these two configurations into the streaming experiment to test how
their timing differences affect correction release.

We next examine whether the single-frame latency advantage translates
into earlier correction release during streaming memory. We combine
GARI ind.128 with 20 retries and $D_{XYZ}$ with 100 retries with
causal window decoding over 512 noisy rounds. Measured service traces
are replayed with FIFO syndrome arrivals every 2.97 ms, followed by
one ideal final extraction. Every window must satisfy both physical
core sectors, and the accumulated correction is checked against the
full syndrome and all 24 logical labels. Rejected windows count as
stream failures. The reported release age includes core/buffer
waiting, computational queuing, and correction processing.

The shorter decoding times of GARI ind.128 translate into earlier
correction release as shown in Fig.~\ref{fig:arbitrary_state}(c). With
the $7/3$ geometry, its p99.9 release age is $36.25$ ms, compared with
$166.29$ ms for $D_{XYZ}$. The corresponding LERs are
$3.24\times10^{-6}$ and $5.38\times10^{-6}$ per round. Queuing
accounts for much of the latency difference: Only $0.40\%$ of
noisy-round arrivals wait for GARI ind.128, compared with $67.88\%$
for $D_{XYZ}$. Their p99.9 computational waiting times are $6.41$
and $136.18$ ms, respectively.

Reducing the core to four rounds releases corrections sooner for both
configurations.
The p99.9 release ages of GARI ind.128 and $D_{XYZ}$ fall to $24.12$ and $67.16$ ms, respectively, and their
queuing tails fall to $3.29$ and $46.27$ ms. The LER point estimates
rise to $4.24\times10^{-6}$ and $7.80\times10^{-6}$, with overlapping
intervals across the two geometries for each configuration.  Thus,
core length remains a useful control over correction delay after the
graph and schedule have been chosen.

Together, the experiments connect BP-SF's parallel retries to its use
in streaming memory. Parallel trials shorten difficult decoding calls,
windowing limits temporal graph growth, and correlation-aware graph
and schedule choices improve the full-frame accuracy--latency
trade-off.  In the tested arrival stream, combining these choices
yields substantially less queuing and earlier correction release.

\endgroup

\section{Conclusion}
\label{sec:conclusion}

\begingroup
We extended oscillation-guided BP-SF from parallel post-processing to
GPU-based windowed and streaming decoding. The GPU implementation
combines an ordinary BP attempt with posterior-initialized, parallel
DMem-BP retries, exposing a direct trade-off between decoding
resources, latency, and accuracy. On the
$\llbracket144,12,12\rrbracket$ code at $p=0.002$, all-concurrent
BP-SF reduces mean and p99.9 latency by $6.28\times$ and
$13.15\times$, respectively, relative to the tested eight-chain
Relay-BP implementation on 100,000 matched samples for which every
method's native initial BP stage fails within 80 iterations. These
conditional timing gains do
not imply uniform accuracy superiority: Relay-BP retains an accuracy
advantage, particularly on the $\llbracket288,12,18\rrbracket$ code.

Fixed-size window decoding extends BP-SF to long memory experiments
without the growing decoding graph of a monolithic approach. In the
144-code fixed-basis experiment, windowed decoding maintains
approximately constant per-round compute time and LER through 512
rounds, where its mean compute time per round is $17.1\times$ lower
than monolithic decoding. Core and buffer selection nevertheless
requires balancing accuracy, decoding frequency, and
correction-release delay, i.e., compute latency alone does not capture
lookahead or queuing costs.

We also evaluated preservation of arbitrary logical states by checking
the complete logical Pauli frame under stochastic Pauli circuit
noise. The memory circuit uses ideal initial and final stabilizer extraction without destructive data-qubit readout, and
the scoring criterion checks both physical sectors and all logical
$X/Z$ labels rather than extrapolating from a single-basis memory
test. In this setting, graph representation and scheduling materially
affect the accuracy--latency trade-off. In the twelve-round 144-code
experiment, GARI with independent-128 layers and 20 BP-SF retries
achieves accuracy comparable at the measured precision to
$D_{XYZ}$ with flooding updates and 100 retries, while reducing mean and p99.9
natural-shot latency by $19.3\times$ and $3.85\times$. Its advantage
also carries into the tested 512-round streaming workload as
substantially less computational queuing.  Together, these results
support co-designing speculative trials, graph structure, and window
geometry, and evaluating both logical correctness and end-to-end
correction release. Extending these gains across code families and
establishing long-run queue stability remain important next steps.
\endgroup

\bibliographystyle{ACM-Reference-Format}
\bibliography{ref}

\appendix
\section{Additional Experimental Details}
\label{app:simulation_details}
\begingroup
\subsection{Full-Frame Protocol and Scoring}
\label{app:full_frame_protocol}
For a syndrome-consistent correction under stochastic Pauli circuit
noise, the residual action on the $k$ encoded qubits is
\begin{equation}
    \label{eq:full_logical_frame}
    E_L = \prod_{j=1}^{k}\bar{X}_j^{a_j}\bar{Z}_j^{b_j},
    \qquad (\bm a,\bm b)\in\mathbb{F}_2^{2k},
\end{equation}
up to an irrelevant global phase. Preserving every encoded input
requires $\bm a=\bm b=\bm0$.

The full-frame circuit has ideal initial $X/Z$ stabilizer extraction,
$R$ noisy extraction rounds, and ideal final $X/Z$ extraction, without
destructive data-qubit readout. Paired Pauli-target annotations in
Stim track each logical $\bar Z_j$ and $\bar X_j$ correlation across
the ideal
boundaries~\cite{gidney2021stim,stim_pauli_observables}. These
simulator annotations record noise-induced changes without physically
measuring non-commuting logical operators. The decoder receives only
the detector syndrome. The logical labels are used for scoring and are
verified against the final data-qubit Pauli frame. All 24 labels of
the 144 code are checked on each sample.

Let $D\bm e=\bm s$ describe the full physical detector model and
$L\bm e$ its logical-frame labels. A memory experiment fails when
\begin{equation}
    \label{eq:full_frame_failure}
    F=\{\text{decoder rejects}\}\;\cup\;
      \{D\hat{\bm e}\ne\bm s\}\;\cup\;
      \{L\hat{\bm e}\ne L\bm e\}.
\end{equation}
All algebra is over $\mathbb F_2$. Any wrong logical label counts as one
failed experiment, and non-convergence is included. This directly scores
both error sectors rather than extrapolating from a single-basis LER.
For streaming experiments, any rejected window fails the stream, and the
final accumulated correction is checked with the same full-frame rule.
All input rounds are retained and there is no monolithic rescue.

\subsection{Decoder Settings}

Windowed and full-frame BP-SF use up to 80 initial min-sum BP
iterations followed, on failure, by DMem-BP trials of up to 60
iterations. Trials initialize posterior memory from the failed initial
attempt, with fixed node-specific strengths
$\gamma_{v_i}\in[-0.24,0.66]$. Up to 50 oscillation-selected
candidates support flip weights 1--10. The window geometry and length
sweeps allocate 10 trials to each weight, giving up to 100 retries per
window, or per memory experiment for the monolithic comparison. Local
window acceptance and candidate selection use the core.  Full-frame
experiments retain the min-sum normalization $1-2^{-t}$ and divide
each configured retry budget equally among the ten flip weights.

Separate full-frame decoding estimates effective $X$ mechanisms using
$D_Z$, then conditions the effective $Z$ priors on the accepted
estimate before decoding $D_X$. The conditional one-bit marginals
account for shared mixed-error mechanisms. Treating the first estimate
as certain and factorizing the second-sector priors are
approximations. If the first sector fails, the second uses its
original marginal priors. The retry budget applies separately to each
failed sector.

For full-frame GARI decoding, we construct $U$ and $V$ from the
joint detector model used in the experiment. The resulting
effective error variables, $\bar{\bm e}_X$ and $\bar{\bm e}_Z$,
are mapped back to corrections with the same detector and
logical effects. BP-SF selects flip candidates only from the
original error-mechanism variables, excluding the auxiliary
variables introduced by GARI. The virtual checks participate
in message passing, but convergence is tested on the measured
$X$- and $Z$-type detector constraints. Both sectors must be
satisfied before a full-frame estimate is accepted.

In the ind.128 schedule, the checks corresponding to measured
detector outcomes are partitioned into layers containing at most
128 checks to increase GPU utilization. Checks within a layer share no variable nodes and
are updated in parallel. The layers are processed sequentially, and
the partition remains fixed throughout decoding. At the beginning
of each iteration, we randomly reorder these physical-check
layers. After processing them, we update the two virtual-check
layers in the fixed order $U$ followed by $V$. One iteration
therefore consists of a complete sweep through the physical
and virtual layers. The limit of 128 applies only to the
physical-check layers. This is different from the original GARI \cite{maan2025decoding}, where physical check updates are performed in a serial manner (i.e., each layer has only one check). 
\subsection{Sampling and Timing}

Figure~\ref{fig:complexity_growth}(c) compares decoding latency on
100,000 matched samples for which every tested method's native initial
BP stage fails to converge within 80 iterations. This common condition
includes ordinary BP in both BP-SF schedules, Relay-BP's initial
memory-BP stage, and the BP stage of BP80-OSD10. Selection uses initial
convergence only, independent of subsequent retry success or logical
correctness. Each sample is decoded once per configuration.
Timing covers the complete scalar call, including initial BP and
subsequent processing, and excludes sampling, scoring, construction,
and warm-up. All runs use the same RTX 5060 Ti under exclusive
allocations, with rotated configuration order.

The 128-round geometry sweep uses matched syndrome samples over the
common prefixes of the sample sets. Each setting is evaluated in
blocks of 4,096 experiments until at least 100 failures are observed,
requiring 16,384--86,016 experiments. A separate set of 2,048 matched
streams per setting supplies scalar timings. An exploratory paired
comparison of $7/3$ and $10/3$ on their common samples gives a nominal
exact McNemar $p=0.085$, which is descriptive under the error-count
stopping rule.  The geometry sweep is independent of the length
sweep. For the latter, LER samples are matched between decoders and
collected in blocks until at least 100 failures are observed for each
decoder. Timings use 2,048 matched memory experiments per
length. Decoder construction and warm-up are excluded, and static
per-window observable matrices are precomputed.

In the twelve-round full-frame accuracy sweep, separate decoding has
at least 100 failures per point, GARI ind.128 has 30, and $D_{XYZ}$
ind.128 has 20. $D_{XYZ}$ with flooding updates at 100, 300, and 1,000 retries reaches
the 4,194,304-sample cap with 32, 15, and 17 failures,
respectively. The nominal 95\% LER intervals for GARI ind.128 with 20
retries and $D_{XYZ}$ with 100 retries are
$[5.57,11.35]\times10^{-7}$ and $[4.50,8.98]\times10^{-7}$. Scalar
timings use 524,288 matched, unfiltered samples per configuration,
including failures, and exclude circuit sampling, scoring, graph
construction, and warm-up.

The full-frame streaming experiment collects 50 failures per
configuration for LER estimation. Separately, 4,096 matched natural
streams per setting provide measured service traces for FIFO arrival
replay. Arrival spacing is 2.97 ms, including one additional arriving
perfect-final extraction.  While graph construction is excluded from
timing, per-arrival processing and final flush are
included. Computational waiting excludes intentional core/buffer
lookahead, while release age includes both. Timing intervals resample
whole streams to preserve any within-stream dependence. These traces
characterize the tested 512-round workload.
\par
\endgroup

\end{document}